# CellProfiler plugins - an easy image analysis platform integration for containers and Python tools

Erin Weisbart[1]*, Callum Tromans-Coia[1]*, Barbara Diaz-Rohrer[1]*, David R Stirling[1]*, Fernanda Garcia-Fossa[1,2]*, Rebecca A Senft[1]*, Mark C Hiner[3], Marcelo B de Jesus[2], Kevin W Eliceiri[3], Beth A Cimini[1+]

[1]=Broad Institute of MIT and Harvard, Cambridge MA, USA
[2]=Department of Biochemistry and Tissue Biology, Institute of Biology, University of Campinas, Campinas, São Paulo, Brazil
[3]=University of Wisconsin-Madison, Madison, WI, USA

[+]=To whom correspondence should be addressed; Contact details:

Dr. Beth Cimini,
Imaging Platform,
Broad Institute,
415 Main St,
Cambridge, MA 02142
Email: bcimini@broadinstitute.org
Phone: 617-714-7000

## Abstract

CellProfiler is a widely used software for creating reproducible, reusable image analysis workflows without needing to code. In addition to the >90 modules that make up the main CellProfiler program, CellProfiler has a plugins system that allows for the creation of new modules which integrate with other Python tools or tools that are packaged in software containers. The CellProfiler-plugins repository contains a number of these CellProfiler modules, especially modules that are experimental and/or dependency-heavy. Here, we present an upgraded CellProfiler-plugins repository, an example of accessing containerized tools, improved documentation, and added citation/reference tools to facilitate the use and contribution of the community.

# Practitioner points

1. CellProfiler plugins expand the kinds of reproducible analyses that can be performed inside CellProfiler.
2. Plugins allow for integration of niche, experimental, or cross-language tools into a central image processing workflow.
3. Containerization is particularly useful in allowing access to tools written in different languages and/or with conflicting dependencies.

# Introduction

Bioimage analysis has become increasingly popular over the last 30 years, with approximately ten times as many citations per year as 20 years ago and three times as many as a decade ago (Fig 1). The number of open-source image analysis tools has both driven and been driven by this increase - since the Scientific Community Image Forum (forum.image.sc)[1] was formed in 2018, the number of participating software tools has grown from 2 to 56. A recent overview of open source image analysis tools or platforms identified 82 tools, platforms, or languages[2], and the BioImage Informatics Index (BIII)[3] lists 1,388 available tools, components, and workflows.

While too many tools is certainly preferable to too few, this increase in software tools and methods means users must wade through an increasingly large tool ecosystem to find the best tool for their needs, which can be especially overwhelming for non-expert users. Deep-learning based tools, also booming in popularity, often have particularly complicated installations. A recent analysis of posts at forum.image.sc implies that possibly as many as one in ten threads is started due to installation issues and as many as one in five touch on installation at some point[4].

Since image analyses rarely are composed of single steps[5], in addition to tools that specialize in performing a single task well, the bioimage analysis space consists of a number of generalist "no-code" (or "code-optional") software workflow platforms that collect other tools and/or help create workflows, including Fiji[6], Icy[7], KNIME[8], and CellProfiler[9,10]. Since their role is at least in part to bring groups of tools together, such workflow platforms typically accept outside contributions or "plugins" created by other groups. It is worth noting that in a recent survey about image analysis in the life and physical sciences, tool users frequently mentioned both "platforms" and "plugins" when asked about what tool developers could do to make user experiences easier[11].

Of the four tools listed above, Fiji, Icy, and KNIME are all implemented in Java; CellProfiler is the sole tool written in Python. Though Java has long been dominant in bioimage analysis due to tools like ImageJ[12,13], Python-based tools are becoming increasingly popular, partially though not exclusively due to Python's dominance in deep learning. The recent creation of the PyImageJ[14] library helps integrate Java- and Python- tools in code-friendly spaces like Jupyter notebooks[15] and napari[16]; nevertheless, a Python-based image analysis workflow tool is an important part of the bioimage analysis ecosystem. As such, while CellProfiler has for a number of years allowed users to implement their own custom plugins[9,10], we herein report on efforts to make CellProfiler-plugins easier to find, install, develop, and use.

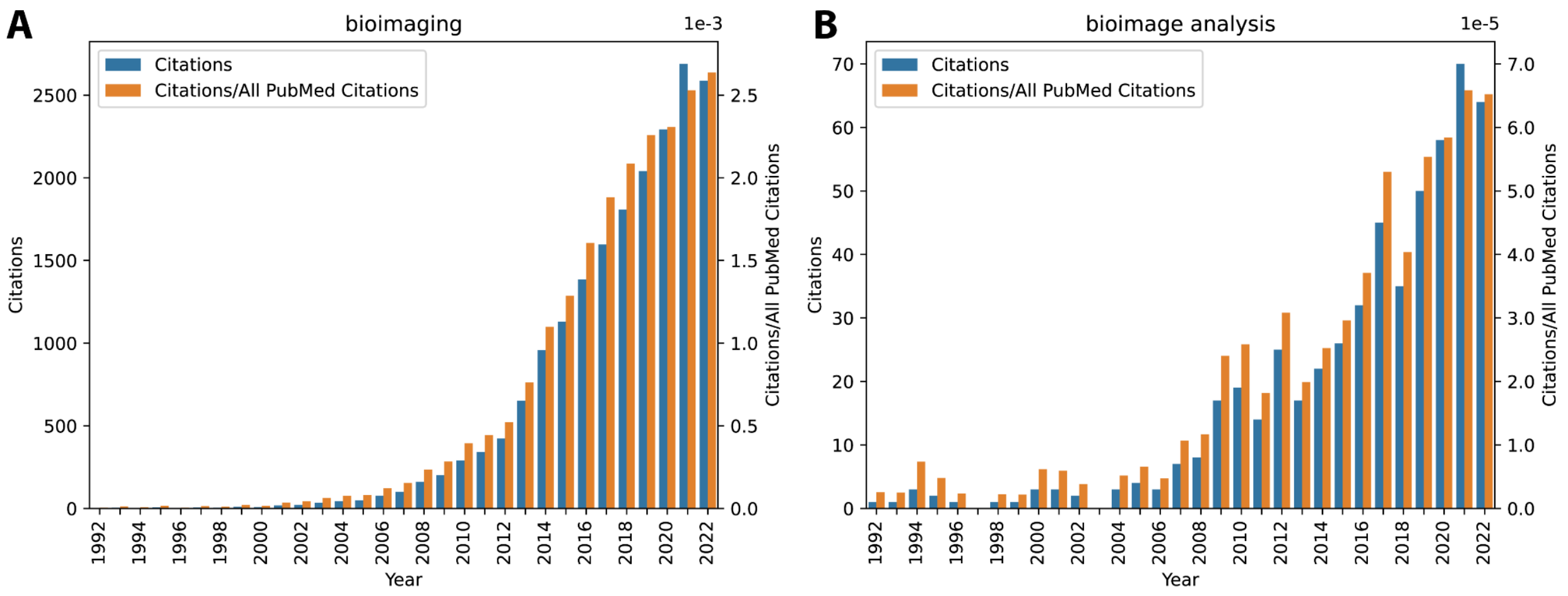

**Figure 1:** PubMed search results for "bioimaging" (A) and "bioimage analysis" (B) from 1992-2022, plotted as absolute citation counts (blue) and citation counts normalized to all PubMed citations (orange). Increase in both terms is substantial over this time window.

# Results

## ***What are CellProfiler plugins?***

CellProfiler enables the flexible creation of image analysis workflows by providing almost 100 modules that each perform an image or object processing function that are arranged and configured by a researcher into a pipeline specific to their analysis task. Plugins are modules that extend CellProfiler's capabilities but are not installed in CellProfiler by default. They allow CellProfiler to be customized and expanded beyond the capabilities of a single development team, and provide a platform for other developers to share their work with the community, making CellProfiler more extensible and collaborative. The CellProfiler-plugins repository stores and shares these CellProfiler modules. Version 1.0.0 of the repository contains 15 plugins (Table 1). Decisions about incorporating plugins into the main program are frequently revisited by the CellProfiler development team in consultation with plugin creators. The two primary, non-mutually-exclusive reasons a module might be distributed as a plugin are narrow audience suitability and library dependencies.

*Modules not suitable for a majority of CellProfiler users*: The existing CellProfiler application contains more than 90 individual modules, and user feedback has consistently shown that the program's complexity is a hurdle to adoption. By consciously including only new modules with a high threshold for reuse, we slow the growth of CellProfiler's complexity; thus, modules such as CallBarcodes, which decodes one-hot-exponentially-multiplexed barcodes such as in in situ sequencing[17] and seqFISH[18], are typically designated as plugins because, in our experience, only a small fraction of CellProfiler users are performing this type of experiment. This class also includes modules we feel have a high potential for accidental misuse: histogram equalization and normalization (performed in HistogramEqualization and HistogramMatching modules) are both image manipulations that are undeniably useful in certain contexts but can introduce hard-to-trace effects if used before image quantification. Thus, these modules are only available to users who seek them out.

*Modules that require external dependencies:* The existing CellProfiler application contains a number of popular scientific Python libraries like NumPy[19], SciPy[20], scikit-image[21], and scikit-learn[22]. It does not, however, contain dataframe handling tools such as pandas[23] nor libraries typically used for deep learning such as TensorFlow[24] or PyTorch[25] due to added complexity for users and to keep the CellProfiler application at a reasonable file size. If the libraries needed to support a particular module are not currently contained in CellProfiler, the module must be used as a plugin, either by using specially-installed versions of CellProfiler or by using software containers (see below). If the module seems to be suitable for the majority of CellProfiler users, the dependencies may be added in future versions of CellProfiler, subject to considerations such as library size, compatibility with existing libraries, and licensing constraints[26] .

| **Plugins with no extra dependencies** | |
|---|---|
| **Module Name** | **Module Purpose** |
| CalculateMoments | CalculateMoments extracts moments statistics from a given distribution of pixel values. |
| CallBarcodes | CallBarcodes is used for assigning a barcode to an object based on the channel with the strongest intensity for a given number of cycles. It is used for optical sequencing by synthesis (SBS). |
| CompensateColors | CompensateColors determines how much signal in any given channel is because of bleed-through from another channel and removes the bleed-through. It can be performed across an image or masked to objects and provides a number of preprocessing and rescaling options to allow for troubleshooting if input image intensities are not well matched. |
| DistanceTransform | DistanceTransform computes the distance transform of a binary image. The distance of each foreground pixel is computed to the nearest background pixel and the resulting image is then scaled so that the largest distance is 1. |
| EnhancedMeasureTexture | EnhancedMeasureTexture measures the degree and nature of textures within an image or objects in a more comprehensive/tuneable manner than the MeasureTexture module native to CellProfiler. |
| HistogramEqualization | HistogramEqualization increases the global contrast of a low-contrast image or volume. Histogram equalization redistributes intensities to utilize the full range of intensities, such that the most common frequencies are more distinct. This module can perform either global or local histogram equalization. |
| HistogramMatching | HistogramMatching manipulates the pixel intensity values of an input image and matches them to the histogram of a reference image. It can be used as a way to normalize intensities across different 2D or 3D images or different frames of the same 3D image. It allows you to choose which frame to use as the reference. |
| PixelShuffle | PixelShuffle takes the intensity of each pixel in an image and randomly shuffles its position. |
| Predict | Predict allows you to use an ilastik pixel classifier to generate a probability image. CellProfiler supports two types of ilastik projects: Pixel Classification and Autocontext (2-stage). |
| VarianceTransform | VarianceTransform allows you to calculate the variance of an image, using a determined window size. It also has the option to find the optimal window size from a predetermined range to obtain the maximum variance of an image. |
| **Plugins with extra dependencies** | |
| RunCellpose | RunCellpose allows you to run Cellpose[27] within CellProfiler. Cellpose is a generalist machine-learning algorithm for cellular segmentation and is a great starting point for segmenting non-round cells. |
| RunDeepProfiler | RunDeepProfiler allows you to create DeepProfiler[28] features within CellProfiler. DeepProfiler is a comprehensive suite of tools designed to leverage deep learning techniques for the analysis of imaging data in high-throughput biological experiments. |
| RunImageJScript | RunImageJScript allows you to run any supported ImageJ script *directly* within CellProfiler using PyImageJ[14]. It is significantly more performant than RunImageJMacro, and is also less likely to leave behind temporary files. |
| RunOmnipose | RunOmnipose allows you to run Omnipose[29] within CellProfiler. Omnipose is a general image segmentation tool that builds on Cellpose. |
| RunStardist | RunStarDist allows you to run StarDist[30] within CellProfiler. StarDist is a machine-learning algorithm for object detection with star-convex shapes making it best suited for nuclei or round-ish cells. You can use pre-trained StarDist models or your custom model with this plugin. |

**Table 1**: CellProfiler Plugins in plugins repository version 1.0.0

***Using CellProfiler plugins***

Plugins can be used by simply downloading the entire plugins repository (or any individual plugin) from the CellProfiler-plugins GitHub repository at https://github.com/CellProfiler/CellProfiler-plugins. In the CellProfiler "preferences" menu, the user sets the CellProfiler plugins directory to the folder containing the plugin(s) they wish to use. When CellProfiler is next opened, it will attempt to load all plugins found in that folder; note that it will not recursively search subfolders. Successfully loaded plugins will be available in the “Add Modules” menu alongside CellProfiler native modules. Plugins relying only on CellProfiler-provided libraries (see Table 1) can be loaded in CellProfiler whether the CellProfiler application was downloaded from the CellProfiler.org website (sometimes referred to as installing "built") or whether CellProfiler was installed in Python on the user's computer (sometimes referred to as “from source”).

When the built CellProfiler application does not contain all of a plugin’s dependencies, additional installation of these dependencies or an alternative approach that circumvents local installation requirements must be used in order for the plugins to work; three methods are described in the paragraphs following. Some of the most impactful plugins add new dependencies: bringing new functionality to CellProfiler and bridging gaps in the core library (e.g. modules with Deep Learning functionality such as RunCellpose or RunStarDist), even crossing boundaries to other languages and tools (e.g. RunImageJScript). As seen in Figure 2, Cellpose[27] is a highly-requested plugin for CellProfiler, due to its easy-to-tune segmentation networks.

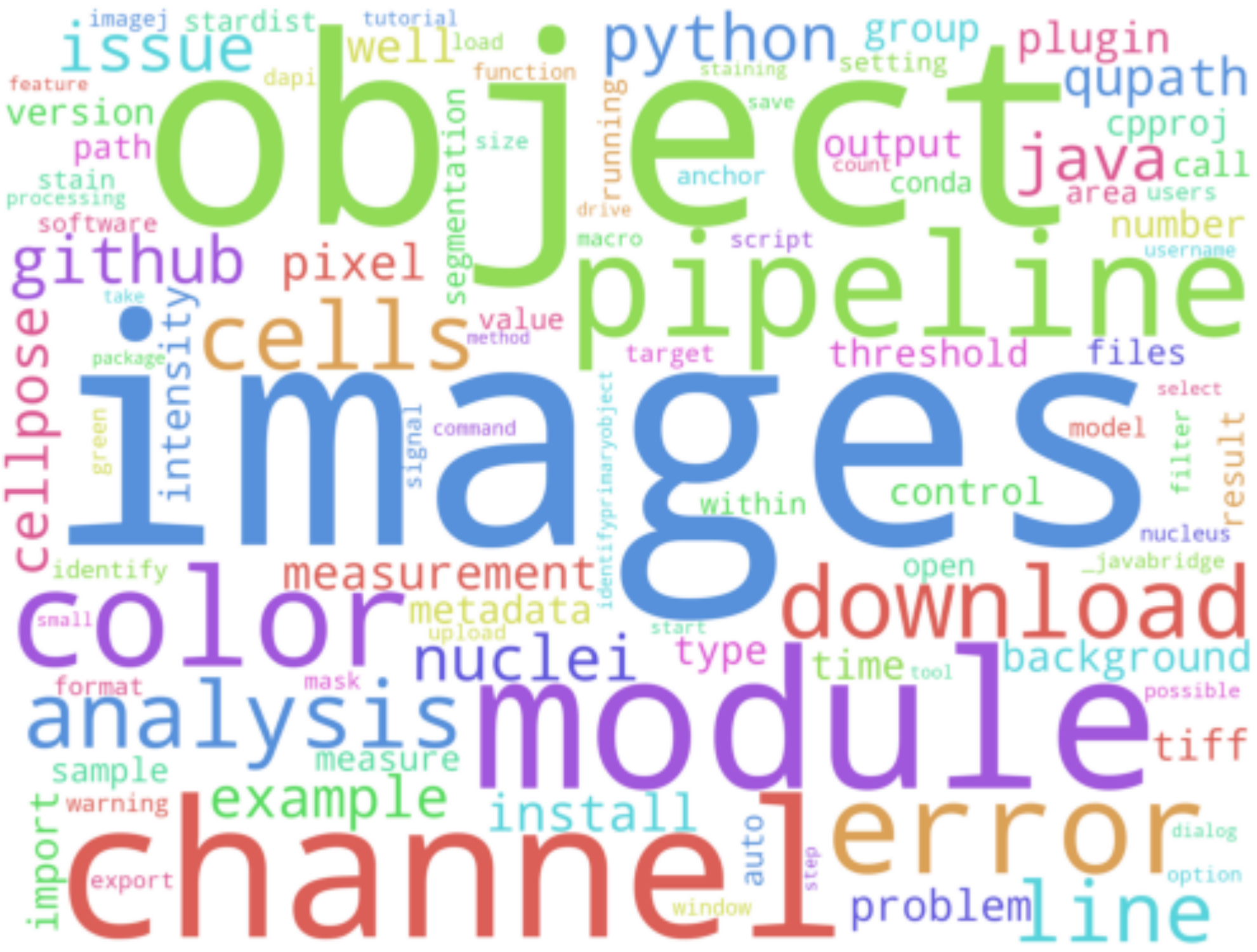

**Figure 2:** Wordcloud of terms used in ~200 forum.image.sc posts tagged with "cellprofiler" from February to May of 2023. "plugin" (top right), "install" (bottom center), "StarDist" (top left), and "cellpose" (center left) are all visible.

*Installation via Python:* CellProfiler can be installed in a system's local Python installation by cloning from GitHub or by directly installing from the Python Package Index (PyPI) in Python versions 3.8 or 3.9. Users may find dependency management easiest if they use Python virtual environment tools such as pyenv[31], or Anaconda[32] and related tools such as mamba[33]. Users can then add additional libraries (as laid out by the plugin author) via pip or conda per their package manager's instructions, though problems may arise when plugins have conflicting dependencies (e.g. Plugin 1 requires a library to be at version 1.5, but Plugin 2 only works with that library at version 2.0). Solving such dependency conflicts typically requires manual review of the dependencies; help is available in the plugins documentation as well as on the image.sc forum, and we provide a setup file in the CellProfiler-plugins repository to help ease pip installation of CellProfiler and each plugin's dependencies. This is the method that we have historically recommended, though it can be admittedly challenging for non-developer users.

*Installation of Python dependencies into a built application:* A developer who wishes to add support for a particular plugin to CellProfiler could add their dependencies directly to the built CellProfiler application and then distribute the modified application, simplifying plugin use for non-developer users. Further explanation is given in the plugins documentation at https://plugins.cellprofiler.org. Since the extra dependencies will be operating system dependent, and since the source of such a modified version should be extremely well trusted before a user attempts to use it, this approach does not scale widely, but could be useful in the context of an individual laboratory or core facility.

*Access of external dependencies via containers:* Since recent analyses suggest users who spend more of their time on "imaging" than "image analysis" self-report lower computational comfort than those who spend more time on image analysis[11], and installation issues can be a particular pain point in working with bioimage analysis software[4], solutions that require Python installation may exclude a sizable fraction of the users of bioimage analysis tools, especially beginning users. Software containers such as Docker[34] and Singularity[35] are increasingly used to distribute open-source software tools; the Biocontainers repository[36] contains more than 1,000 software tools packaged into containers. Containers simplify dependency management (since packages are installed once and only once during creation of the container) and especially version conflicts between packages (since each tool in a workflow can live in a separate container or application). A built copy of CellProfiler that can access containers can therefore, in theory, access any other software that has been containerized. With this in mind, we have recently converted the RunCellpose module to optionally use a Cellpose Docker container so that Python and Cellpose installation are no longer required, which makes our RunCellpose plugin accessible to even the most computationally novice users as the only additional installation it requires is the one-click download of Docker. We are currently working to bring the optional use of Docker to other plugins which are compatible with CellProfiler 4 (the current version), and to modules packaged with CellProfiler 5.

***Creating new CellProfiler plugins***

For developers wishing to create new CellProfiler plugins, a number of resources are available. Module templates are available in the main CellProfiler repository at https://github.com/CellProfiler/CellProfiler/tree/master/cellprofiler/modules/plugins; the plugin templates walk developers through the various steps required to create several classes of CellProfiler modules (such as "image processing" and "measurement"). Additional documentation about plugins and the plugin repository submission process is available at https://plugins.cellprofiler.org. Finally, a video in the NEUBIAS academy YouTube series[37] walks developers through the process of designing and creating a new plugin, including incorporating module settings, design decisions, and how to incorporate the actual run code into the module, which is typically the simplest part of writing a new plugin.

While the above approaches are suitable for addition of everything from simple library calls to large external tools to CellProfiler, developers of the latter may reasonably be concerned that by distributing their tool inside a CellProfiler plugin, the end user may not fully appreciate that it is an independent piece of software and forget to credit and/or cite it. To alleviate these concerns, CellProfiler has added a "doi" property to modules, and added Digital Object Identifier (DOI) citations to all existing plugins relying on external libraries. These DOIs feed into a citation generator tool in CellProfiler that allows users to easily export a list of citations required by their pipeline. This functionality is currently available in CellProfiler's source code, and will be widely available in the upcoming CellProfiler 5 release.

# Discussion

In the 17 years since the original CellProfiler paper[38] was published, CellProfiler has been cited or mentioned in more than 15,000 academic papers. After a code-free pipeline creation process that the user can tune to their own data, it is designed to multiprocess small-to-medium datasets locally on the user's computer or, with a few clicks, it can export files allowing a locally-tuned pipeline to be used on clusters[38] or the cloud[9]. These abilities have earned CellProfiler a reputation as a friendly and easy-to-use platform[39]. In contrast to CellProfiler's early days, however, we now live in a scientific world where new bioimage analysis libraries and tools become available weekly or even daily. We argue that having central workflow platforms that make it easy for developers to share their work and users to easily try it facilitates adoption and knowledge sharing, accelerating the field as a whole.

CellProfiler is certainly not the only such workflow platform - Fiji notably contains more than 10,000 externally contributed plugins, and other tools exist in this space, some of which require minimal or no code (Icy, KNIME) and some of which require code for many or most applications (Jupyter, napari). Within this ecosystem, CellProfiler specializes in shareable, code-free, linear image analysis workflows; users should choose the platform that best fits their needs. Efforts to make multiple deep learning models available, such as DeepImageJ[40] and the Bioimage Model Zoo[41], while not workflow tools, also help users find image analysis solutions, and to try a number of different deep learning tools side-by-side. However, CellProfiler's module templates and ability to use containerized tools help ease the lift for developers in creating easily-disseminated plugins, and contribution to a centralized repository helps both developers and users with discovery of new tools. These changes, along with improved ease of citation, we believe make CellProfiler's plugin system a valuable part of the image analysis workflow ecosystem.

# Conclusion

Bioimage analysis is now a firmly established part of the bioimaging universe; qualitative conclusions drawn from human observations are slowly but steadily being replaced by quantitative measurements of particular image or object properties. This replacement process can be accelerated by giving software users friendly, interactive workflow tools that give them beginning-to-end solutions for their image analysis tasks. Plugins play an important role in this process as they allow users to perform specific but necessary steps to customize and extend their analyses. Plugin creation allows for rapid and straightforward integration of new tools, components, and approaches, while allowing users continuity using workflow tools they are already familiar with. We hope the improvement of the CellProfiler-plugins repository encourages developers to contribute plugins, and encourages users to integrate these plugins into their CellProfiler pipelines. These approaches will lead to faster workflow creation and more reproducible workflows for all those using these tools to solve their image analysis problems.

# Materials and Methods

Current modules in the CellProfiler plugins repository are written for Python[42] 3.8 and designed to be used with CellProfiler 4[10] - the repository also contains unmaintained plugins designed for CellProfiler 2[43] and CellProfiler 3[9]. Dependencies for each plugin are indicated in the repository's setup.py file.

Figures in this paper were generated in Jupyter notebook[15] 6.4.12 running Python 3.8.16 using the matplotlib[44] 3.7.1, pandas[23] 1.5.2, seaborn [45] 0.12.2, and wordcloud[46] 1.9.2 libraries.

# Funding

The work was supported by the Center for Open Bioimage Analysis (COBA) funded by National Institute of General Medical Sciences P41 GM135019 awarded to BAC and KWE. The work was also supported by grant number 2020-225720 to BAC from the Chan Zuckerberg Initiative DAF, an advised fund of Silicon Valley Community Foundation. Funding was also provided by the São Paulo Research Foundation (FAPESP) #2022/01483-4, #2019/24033-1, and #2020/01218-3. The funders had no role in study design, data collection and analysis, decision to publish, or preparation of the manuscript.

# Competing interests

The authors declare that there are no competing interests associated with the manuscript.

# Acknowledgements

The authors would like to thank (in approximate chronological order) Lee Kamentsky, Mark Bray, Allen Goodman, Madison Bowden, Claire McQuin, Dan Ruderman, Jane Hung, Kyle Karhohs, Duaa Ali, Anne Carpenter, Genevieve Buckley, Tim Becker, Curtis Rueden, Carla Iriberry, Filip Mroz, Volker Hilsenstein, Christian Clauss, Niklas Rindtorff, Nick Whalen, Ioannis K. Moutsatsos, Pearl Ryder, Nesta Bentum, Alice Lucas, Mario Cruz, Jenna Tomkinson, Kevin J. Cutler, Dominic Chomchai for submitting plugins and bugfixes to the plugins repository. We are also grateful to the authors of CellProfiler plugins that live elsewhere, as well as the creators of Cellpose, StarDist, Omnipose, DeepProfiler, and PyImageJ for the creation of excellent tools that help the bioimage analysis community.

Co-authors BDR, CTC, DRS, EW, FG-F, and RAS all contributed equally to the work presented in this manuscript, and each has the right to list themselves first in author order on their CVs.

# Code availability statement

The CellProfiler Plugins repository is available at https://github.com/CellProfiler/CellProfiler-plugins and on Zenodo at DOI https://doi.org/10.5281/zenodo.8000109 . Additional code and data used in analyses reported here are available at https://github.com/COBA-NIH/CellProfilerPluginsPaper . Extended CellProfiler Plugins documentation is available at https://plugins.cellprofiler.org/ .